\documentclass[pra ,superscriptaddress, preprintnumbers, amsmath, amssymb]{revtex4}

\usepackage{epsfig}
\usepackage{amsmath}
\usepackage{amssymb}
\usepackage{amsfonts}
\usepackage{mathptmx}
\usepackage{eucal}
\usepackage{bm}
\usepackage{color}
\usepackage[colorlinks,linkcolor=blue,citecolor=blue]{hyperref}
\usepackage{epstopdf}
\usepackage{bbold}
\usepackage{url}
\usepackage{mathtools}
\usepackage{dsfont}
\usepackage{tensor}
\usepackage{dsfont}
\usepackage{bbding}
\usepackage{pifont}
\usepackage{wasysym}
\usepackage{xspace}
\usepackage{enumerate}

\newcommand{\Trace}{{\rm Tr}}

\def \ket#1{\mathinner{|{#1}\rangle}}

\newcommand{\matrixel}[3]{{\mathinner{\langle{#1}| {#2} | {#3}\rangle}} }

\newcommand{\GFunc}{\mathcal{G}_\lambda}

\renewcommand{\email}[1]{\thanks{Email address: #1}}

\def \QNDM{QNDM}
\def \DM{DM}

\begin{document}

\title{A Novel Approach to Reduce Derivative Costs in Variational Quantum Algorithms}

\author{G. Minuto}
\email{giovanni.minuto@uniroma1.it}
\affiliation{Dept. of Informatics, Bioengineering, Robotics, and Systems Engineering (DIBRIS), Polytechnic School of Genoa University, Genova, Italy}
\affiliation{INFN - Sezione di Genova, via Dodecaneso 33, I-16146, Genova, Italy}

\author{D. Melegari}
\email{dmelegari@infn.it}
\affiliation{Dipartimento di Fisica, Universit\`a di Genova, via Dodecaneso 33, I-16146, Genova, Italy}
\affiliation{INFN - Sezione di Genova, via Dodecaneso 33, I-16146, Genova, Italy}

\author{S. Caletti}
\email{scaletti@phys.ethz.ch}
\affiliation{Institute for Theoretical Physics, ETH, CH-8093 Z\"urich, Switzerland}

\author{P. Solinas}
\email{paolo.solinas@unige.it}
\affiliation{Dipartimento di Fisica, Universit\`a di Genova, via Dodecaneso 33, I-16146, Genova, Italy}
\affiliation{INFN - Sezione di Genova, via Dodecaneso 33, I-16146, Genova, Italy}

\date{\today}

\begin{abstract}
We present a detailed numerical study of an alternative approach, named Quantum Non-Demolition Measurement (\QNDM) \cite{Solinas_2023}, to efficiently estimate the gradients or the Hessians of a quantum observable.
This is a key step and a resource-demanding task when we want to minimize the cost function associated with a quantum observable.
In our detailed analysis, we account for all the resources needed to implement the \QNDM~approach with a fixed accuracy and compare them to the current state-of-the-art method \cite{Mari2021, SchuldPRA2019, Cerezo2021}.
We find that the \QNDM~approach is more efficient, i.e. it needs fewer resources, in evaluating the derivatives of a cost function. These advantages are already clear in small dimensional systems and are likely to increase for practical implementations and more realistic situations. A significant outcome of our study is the implementation of the \QNDM~method in Python, provided in the supplementary material \cite{qndm_gradient}. Given that most Variational Quantum Algorithms can be formulated within this framework, our results can have significant implications in quantum optimization algorithms and make the \QNDM~approach a valuable alternative to implement Variational Quantum Algorithms on near-term quantum computers.
\end{abstract}

\maketitle

\section{Introduction}
\label{sec:intro}
Complex optimization problems, such as drug, molecular, and material design, are some of the research fields in which quantum computers are likely to have an impact in the mid-term timescale \cite{Srinivasan,Moll_2018,D0SC05718E}. 
These problems share a common structure: given a cost function that represents our physical quantity of interest, we would like to find its minimum.
The scheme to approach these problems in a quantum computer is a hybrid quantum-classical one, where Variational Quantum Algorithms (VQA) \cite{McClean_2016} can be naturally implemented.
The picture is the following: quantum computers are used to calculate the gradient of the cost function evaluated at a certain point in the parameter space. This can be done with different techniques but is usually done by evaluating the cost function in two points that are close in the parameter space and exploiting the parameter-shift rule to compute the derivative \cite{Mari2021}.
Then, this information is fed into a classical computer, which calculates how to adjust the parameters of the quantum circuit in order to reduce the cost function.
By iterating this procedure, as in variational problems in physics, the quantum circuit should converge toward the configuration that generates one of the states minimizing the cost function (either a local or a global minimum).

The classical optimization procedures are well-known and established \cite{Boyd_2004,adby2013introduction} and they usually exploit information on the derivatives (gradient) of the cost function to update the parameters.
At the quantum level, the way to extract the desired information from a quantum system is still debated and open \cite{Banchi2021,Wierichs_2022}.
Indeed, the derivative cannot be directly associated with a Hermitian operator \cite{solinas2015fulldistribution, solinas2016probing} and, therefore, there is no unique way to measure it.
Still, this is a crucial step since it usually causes a bottleneck in computation with quantum computers.
The most straightforward and used approach to obtain the gradient of the cost function is to run a quantum circuit and measure a quantum observable at a certain point in the parameter space. Then, we repeat the procedure changing the circuit parameters by a little along each direction, separately \cite{Mari2021, SchuldPRA2019, Cerezo2021}.
We call this method, the Direct Measurements (\DM) approach and, for every point in the parameter space, it requires two observable measurements to determine each component of the gradient. 

In this article, we analyze a novel method to compute the derivatives of the cost function implemented on the quantum circuit. This is called the Quantum Non-Demolition Measurement (\QNDM) approach and it was first theoretically described in Ref. \cite{Solinas_2023}. 
The main idea is to extract the information about the derivative with a single measurement of a quantum observable.
This is obtained by coupling a quantum {\it detector} to the original quantum system so that the information about the value of the derivative of the cost function is stored in the phase of the detector. Finally, the phase is measured to extract this information \cite{Solinas_2023}.

In the following, we present a detailed comparison between the two methods. We consider the evaluation of the cost function at a fixed accuracy, and we extract information about the repetitions of the circuit evaluation required. We do that by estimating the statistical error (Mean Squared Error) in the two approaches, and we calculate the total resource cost in both cases.
This theoretical analysis is followed by full computational simulations, in which we compare the total resources needed in the two approaches to estimate the derivative of the cost function directly from their implementations.

We find that, already for cases with limited complexity, the \QNDM~approach has a substantial advantage over the \DM~ones. The resource-saving will increase for more practical and interesting problems of intermediate dimension.
%
We observe that there are other methods that rely on measuring an additional detector, such as the approach proposed in  \cite{guerreschi2017practicaloptimizationhybridquantumclassical}. However, the computational cost of obtaining the derivative with this method is comparable to that of the \DM~approach.

The paper is structured as follows.
In Sec. \ref{sec:der}, we recall the basic ideas of VQA and the two approaches to derivative evaluation.
Section \ref{sec:bv} is devoted to the analysis of the statistical errors and the resource cost. 
Section \ref{sec:cost_sim} presents the numerical comparison, while Sec.\ref{sec:conclusions} contains the conclusions.
The codes to calculate derivatives with \QNDM~used within this article are accessible via GitHub \cite{qndm_gradient}. In Appendix \ref{sec:git} we show how to install the \QNDM~package.


\section{Theoretical Discussion}
\label{sec:der}

\subsection{General Settings}
\label{sec:set}

\begin{figure}[h!]
  \centering
  \includegraphics[width=0.8\linewidth]{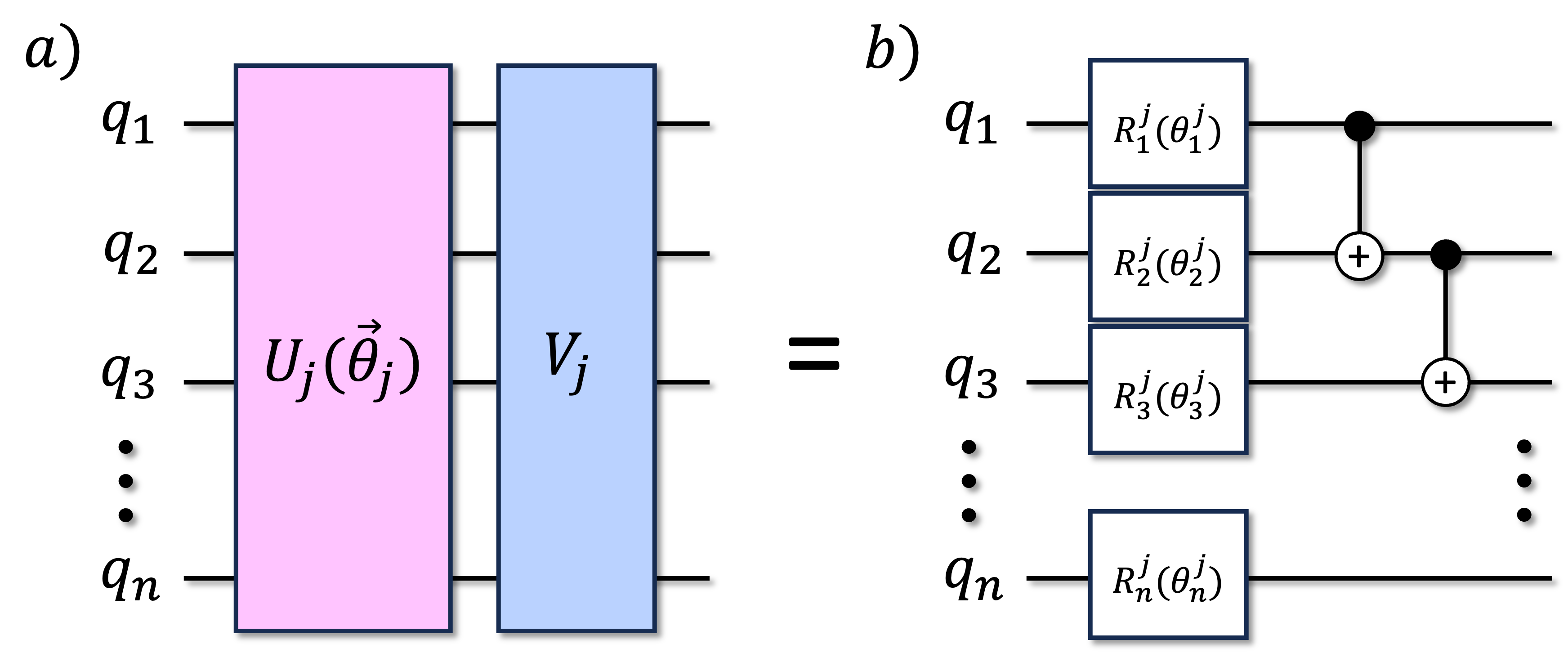}  
  \caption{a) Structure for the $j$-th layer as a sequence of a parameterized layer $U_j(\Vec{\theta}_j)$ and an unparameterized layer $V_j$ for a $n$ qubit system. b) Example of a layer in terms  of single-qubit gates depending on the $\theta_i^j$ parameters and two qubits gates $C_{i-1}NOT_i$.}
  \label{fig:layer}
\end{figure}

We briefly recall the standard implementation of Variational quantum algorithms in a quantum computer (for more details see Refs. \cite{McClean2018, Mari2021, Solinas_2023}).
We consider a system of $n$ qubits initialized in the state $|\psi_0 \rangle = |0\rangle_1\otimes|0\rangle_2 \otimes\ldots \otimes |0\rangle_n \equiv |00...0 \rangle$.
The generic unitary transformation acting on them is denoted $U({\vec \theta})$, where $\vec \theta$ is a vector in the parameter space.
The quantum state obtained by applying the transformation $U({\vec \theta})$ to the initial state of the circuit is given by $|\psi({\vec \theta})\rangle = U({\vec \theta})|\psi_0\rangle$. 
The $U({\vec \theta})$ transformation can be implemented as a sequence of {\it parameterized} unitary transformations $U_j({\theta_j})$ and arbitrary transformations (independents of $\vec \theta$) $V_j$, $j\in[1,m]$, such that
\begin{equation}
 	U({\vec \theta}) = V_m U_m({ \theta_m})....V_2 U_2({ \theta_2}) V_1 U_1({\theta_1})\,.
	\label{eq:U_def}
\end{equation}
We consider $U_j({\theta_j}) = \exp{\{- i \theta_j \hat{H}_j \}}$ and $\hat{H}_j$ is the corresponding generator of the transformation.
The quantum circuit to generate this transformation is represented in Fig.~\ref{fig:layer} a).
Assuming that $\hat{H}_j^2 = \mathds{1}$, we have that 
\begin{equation}
	U_j( \theta_j) = e^{- i \hat{H}_j \theta_j/2} = \cos \frac{\theta_j}{2} \mathds{1} -i \sin \frac{\theta_j}{2} \hat{H}_j.
    \label{eq:rot_par}
\end{equation}
This accounts for a multitude of relevant cases such as when $\hat{H}_j$ is a tensor product of any multi-qubits Pauli matrices \cite{Mari2021}. 
However, to further simplify the discussion and the simulation, in the following, we restrict our attention to the case (presented in Fig. \ref{fig:layer}) in which 
\begin{equation}
    U_j(\vec \theta_j) = R^j_1(\theta^j_1)\dots R^j_n(\theta^j_n)\,,
    \label{eq:rot_layer}
\end{equation}
where each operator $R^j_i(\theta^j_i)$ is a parameterized single qubit gate (see Eq.  (\ref{eq:rot_par})) and $\hat{H}_j$ is a single Pauli matrix. Likewise, for the $V_j$ part of the transformation, which is independent of the parameters, we select the following specific structure
\begin{equation}
    V_j = C_1NOT_2\dots C_{n-1}NOT_n\,,
    \label{eq:ent_layer}
\end{equation}

where $C_{i-1}NOT_i$ is a two-qubit operator with the control on the $i-1$-th qubit and the action on the $i$-th
qubit \cite{nielsen-chuang_book}.
The sequence of $U_j(\vec \theta_j)$ and $V_j$ is usually called {\it layer} and, as shown in Fig. \ref{fig:layer}, is a combination a parametrized and entangling unitary transformations. 
%

In optimization problems, the goal is to find the quantum state $\ket{\psi(\vec\theta)}$ which minimizes the expectation value of a quantum observable $\hat{M}$.
To accomplish this task, we define a cost function $f$ that we want to minimize with respect to the parameters ${\vec \theta}$
\begin{equation}
	f({\vec \theta}) = \matrixel{\psi ({\vec \theta})}{\hat{M}}{\psi ({\vec \theta})} = \matrixel{0}{ U^\dagger({\vec \theta}) \hat{M} U({\vec \theta})}{0}\,.
	\label{eq:f_def}
\end{equation}

In many applications \cite{McArdle2020,TILLY20221}, the observable $\hat{M}$ can be written as a weighted sum of Pauli strings $\hat{P}_i = \prod_j \hat{A}_i^j$, where $\hat{A}_j \in [X_j, Y_j, Z_j, I_j]$ is one of the usual Pauli operators acting on the $j$-th qubit.
With this notation, we have that 
\begin{equation}
    \hat{M} = \sum_{i=1}^J h_i \hat{P}_i
    \label{eq:M}
\end{equation}
where $J$ is the total number of Pauli string composing $\hat{M}$.

We adopt the so-called hybrid quantum-classical optimization procedure described in the introduction \cite{McClean_2016, Cerezo2021}.
The minimization algorithm consists of two steps that must be iterated until the minimum of $f({\vec \theta})$ is reached.
First, with the help of a quantum computer, we evaluate the gradient of $f({\vec \theta})$ in a given point of the parameter space $\vec \theta$. Then, the information about the gradient is fed into a classical computer which, with standard optimization algorithms \cite{kingma2017adam,Kubler2020adaptiveoptimizer, Sweke2020stochasticgradient}, evaluates in which direction of the $\vec \theta$ space we move to reach the minimum of $f$.
In this paper, we focus on the quantum part of the algorithm, that is the calculation of the gradient of $f({\vec \theta})$ done with a quantum computer and we do not discuss the classical one.

To measure the derivative of $f({\vec \theta})$ along the $l$-th direction in the parameter space $\vec \theta$, we measure $f(\vec \theta +s \hat e_l )$ and $f(\vec \theta -s \hat e_l )$ where $\hat e_l$ is the unit vector along the $\theta_{l}$ direction and $s$ is a shift parameter.
Then, we calculate the quantity \cite{Mari2021}
\begin{equation}
	g_{l} = \frac{\partial f({\vec \theta})}{\partial \theta_{l}} = \frac{f(\vec \theta +s \hat e_l ) - f(\vec \theta -s \hat e_l)}{ 2 \sin s }\,.
	\label{eq:f_grad_def}
\end{equation}
Notice that $g_{l}$, unlike finite difference methods, is not an approximation but the exact derivative of $f({\vec \theta})$. By repeating this procedure for all directions $\theta_l$ we obtain the full gradient of $f({\vec \theta})$, which is then fed into the classical optimization algorithm. Setting $s=\pi/2$ yields the parameter shift rule described in Refs. \cite{Li2017,Mitarai2018,SchuldPRA2019,Mitarai2019}.

\subsection{Direct Measurement}
\label{sec:DM}
The standard approach to measure the value of the derivatives of the cost function (\ref{eq:f_grad_def}) is the Direct Measurement (\DM) method \cite{SchuldPRA2019}.
This method consists of measuring separately the average values of the operator $\hat M$ in the two points, i.e., $\vec \theta \pm s \hat e_l$, and then calculating the derivative as in Eq. (\ref{eq:f_grad_def}).
To calculate the values of $f(\vec \theta +s \hat e_l )$, we run the quantum circuit to implement the corresponding $U(\vec \theta +s \hat e_l)$ transformation, and then we perform a projective measurement {\it for each Pauli string} $\hat{P}_i$ that appears in the definition (\ref{eq:M}) of the observable $\hat M$ \cite{McArdle2020, Mari2021}.
For every Pauli string $\hat{P}_i$ in $\hat{M}$, we have to iterate these steps until the statistical accuracy is reached.
The same procedure is then implemented for $f({\vec \theta -s \hat e_l })$ and then, the derivative can be calculated as in Eq. (\ref{eq:f_grad_def}).

\begin{figure}[h!]
  \centering
  \includegraphics[width=0.7\linewidth]{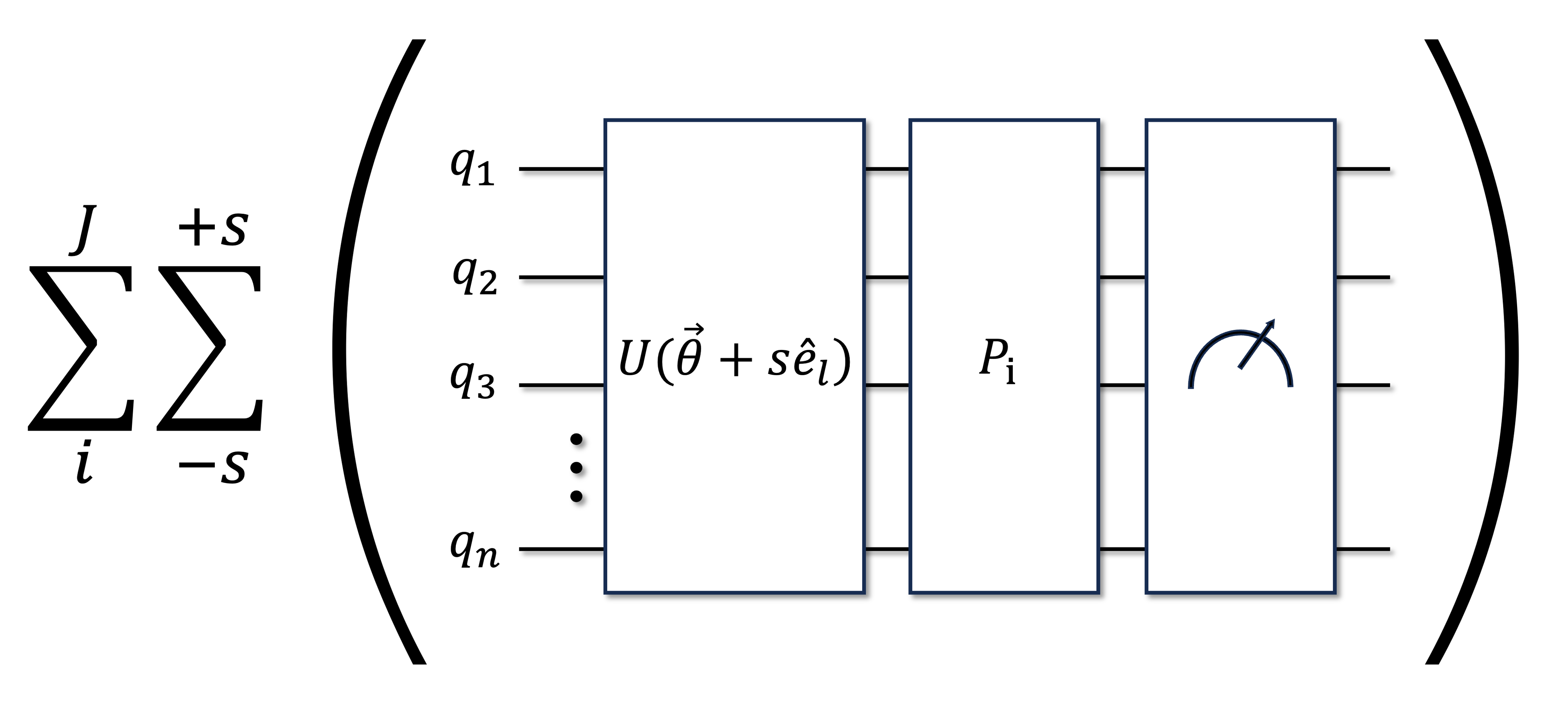}  
  \caption{Pictorial representation for calculation of derivative of the function $f({\vec \theta}) = \matrixel{0}{ U^\dagger({\vec \theta}) \hat{M} U({\vec \theta})}{0}$ with $\hat{M} = \sum_{i=1}^J h_i \hat{P}_i$ using the \DM~protocol, as described in Eq. \ref{eq:DM}. The $n$ qubits circuit is is evolved under the parametric transformation $U(\vec \theta + s\hat{e}_l)$, defined similarly to Eq.\ref{eq:U_def}, followed by a base rotation $P_i$, which represents the $i$-th tensor product of Pauli matrices. A global measurement is then performed on the register, yielding a real-valued output.
  The first summation indicates that the circuit must be executed for each Pauli string ($i = 1, \ldots, J$) in the observable $\hat{M}$. The second summation accounts for the execution of the circuit with different parameter shifts $\pm s$, as shown in Eq.~(\ref{eq:DM}).  
  }
  \label{fig:algo_str_DM}
\end{figure}

Following this procedure, a single derivative $g_l$ in the \DM~approach can be written as
\begin{equation}
    g_l = \sum_i \frac{h_i}{2 \sin s}\Big( \Trace_S \big[\hat{P}_iU^{\dagger}(\vec \theta+s \hat e_l)\rho_s^0U(\vec \theta+s \hat e_l) \big] -\Trace_S \big[\hat{P}_iU^{\dagger}(\vec \theta-s \hat e_l)\rho_s^0 U(\vec \theta-s \hat e_l) \big]\Big)\,,
    \label{eq:DM}
\end{equation}
where $\rho_s^0 =|\psi_0\rangle\langle \psi_0|$, $U(\vec \theta)$ is the operator in Eq. (\ref{eq:U_def}) and $\Trace_S$ denotes the trace over all the qubits. An implementation in terms of quantum circuits is shown in Fig. \ref{fig:algo_str_DM}. 

\subsection{Quantum Non-Demolition Measurement}
\label{sec:QNDM}
In this section, we briefly recall the main characteristics of the \QNDM~method \cite{Solinas_2023}. 
The key idea is to store the information about the derivative of the cost function in the phase of an ancillary qubit named the {\it detector}.
More specifically, by coupling the original system and the detector twice, it is possible to store the information about both $f(\vec \theta +s \hat e_l )$ and $f(\vec \theta -s \hat e_l )$ in the phase of the detector.
Then, the detector is measured to extract information about the derivative \cite{Solinas_2023}.

The advantage of this approach lies in the fact that the couplings between the system and the detector are sequential and the measure on the detector is performed only at the end of the procedure.
Therefore, even if the quantum circuit must be iterated to obtain the desired statistical accuracy, we extract the derivative information by performing just a single measurement on the detector qubit instead of the two needed with the DM approach.
The details of the \QNDM~algorithm can be found in \cite{Solinas_2023}. Here, we summarize only the main steps.

Suppose we are interested in the quantum observable  $\hat{M}$ in Eq. (\ref{eq:M}).
We add an ancillary qubit and implement the system-detector coupling through the operator $U_\pm = \exp\{\pm i \lambda Z_a \otimes \hat{M} \}$ where $Z_a$ is a Pauli operator acting on the ancillary detector and $\lambda$ is the system-detector coupling constant. We notice that if $\hat{M}$ is a complex operator, the implementation of operator $U_\pm$ is, in general, very resource-consuming. This might seem a critical drawback, but the \QNDM~approach offers a clear and elegant way to bypass this problem, instead of implementing the operator $U_\pm = \exp\{\pm i \lambda Z_a \otimes \sum_{i=1}^J h_i \hat{P}_i\}$, we can implement the product of single Pauli string operators $\prod_i^J\exp\{\pm i \lambda Z_a \otimes h_i\hat{P_i} \}$ without relying on the Trotterization approximation \cite{Hatano_2005}, as demonstrated in \cite{Solinas_2023}.
Considering the described quantum circuit, this corresponds to the following transformation (in the full system and detector Hilbert space)
\begin{equation}
	U_{tot} =  e^{ i \lambda Z_a \otimes \hat{M} } U(\vec \theta +s \hat{e}_l ) U^\dagger(\vec \theta -s \hat{e}_l ) e^{-i \lambda Z_a \otimes \hat{M} } U(\vec \theta -s \hat{e}_l )\,.
	\label{eq:U_tot}
\end{equation} 

The initial state for the system plus the detector $\ket{\psi_0}(\ket{0}_D+ \ket{1}_D)/\sqrt{2}$, where a Hadamard gate is applied to the initial state of the detector.
After implementing the transformation (\ref{eq:U_tot}), we measure the accumulated phase $\exp \{i\phi(\lambda)\}$ on the detector degrees of freedom.
It can be shown \cite{Solinas_2023, solinas2015fulldistribution, solinas2016probing, solinas2021, solinas2022} that this is a quasi-characteristic function $\GFunc$
\begin{equation}
	\GFunc \equiv e^{i\phi(\lambda)} =  \frac{\tensor[_D]{\matrixel{0}{\rho_D^f }{1}}{_D}}
	{
	\tensor[_D]{\matrixel{0}{\rho_D^0 }{1}}{_D}
    }\,,
	\label{eq:G_lambda_def}
\end{equation}
where $\ket{0}_D$ and $\ket{1}_D$ are eigenstates of the detector operator $Z_a$, while $\rho_D^0$ and $\rho_D^f$ are the density matrices of the detector before and after the application of $U_{tot}$, respectively.

\begin{figure}[h!]
  \centering
  \includegraphics[width=0.8\linewidth]{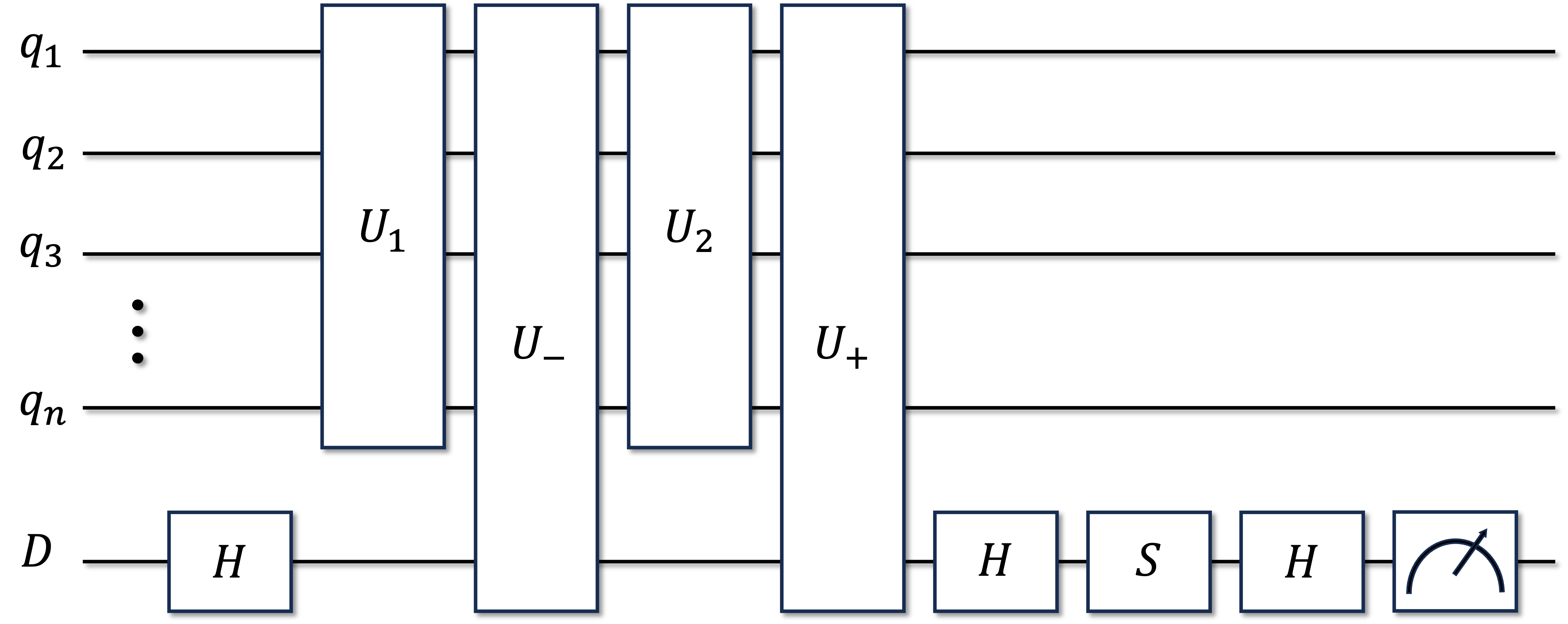}  
  \caption{Pictorial representation of the quantum circuit implementing the \QNDM~protocol. 
  Here, $H$ is the Hadamard gate, $S$ is the phase gate, $U_1 = U({\vec \theta} -s \hat e_l)$, $U_2 = U^{\dagger}({\vec \theta} -s \hat e_l)U({\vec \theta} +s \hat e_l)$ and $U_\pm = \exp\{\pm i \lambda \hat{Z_a} \otimes \hat{M} \}$ is the system-detector coupling operator.}
  \label{fig:algo_str}
\end{figure}

Taking the derivatives of $\GFunc$ with respect to $\lambda$ and evaluating them in $\lambda = 0$, we have access to the information of the different moments of the distribution describing the variation of the cost function \cite{Solinas_2023, solinas2015fulldistribution, solinas2016probing, solinas2021, solinas2022}.
Here, we are interested only in the first moment which directly gives us $g_l$.
Following Ref. \cite{Solinas_2023}, it can be shown that the first derivative of $\GFunc$ reads
\begin{eqnarray}
	-i \partial_\lambda \GFunc \Big |_{\lambda=0} &=& 2 \Trace_S [U^\dagger(\vec \theta +s \hat{e}_l ) \hat{M} U(\vec \theta +s \hat{e}_l ) \rho_S^0 - U^\dagger(\vec \theta -s \hat{e}_l ) \hat{M} U(\vec \theta -s \hat{e}_l )\rho_S^0] \nonumber \\
    &=&  2 \sum_i h_i \Trace_S [U^\dagger(\vec \theta +s \hat{e}_l ) \hat{P_i} U(\vec \theta +s \hat{e}_l ) \rho_S^0 - U^\dagger(\vec \theta -s \hat{e}_l ) \hat{P_i} U(\vec \theta -s \hat{e}_l )\rho_S^0]
    \,,
	\label{eq:first_derivative}
\end{eqnarray}
where $\Trace_S$ denotes the trace over the {\it system} degrees of freedom.
By direct comparison, Eq. (\ref{eq:first_derivative}) is proportional to Eq. (\ref{eq:DM}),
\begin{equation}
    -i \partial_\lambda \GFunc \Big |_{\lambda=0} = 2\sin(s)g_l.
\end{equation}
Thus, both the \QNDM~and the \DM~approach can be used to compute the derivative of $f(\vec \theta)$. 

We note that it is possible to approximate, at the linear regime, the derivative of the quasi-characteristic function to the argument of the accumulated detector phase $\phi(\lambda)$\cite{Solinas_2023}. In practice, this approximation holds if the coefficient in the argument of the system-detector coupling operator, $U_\pm = \exp\{\pm i \lambda \hat{Z_a} \otimes \hat{M} \}$, is small. This condition is satisfied for sufficiently small $\lambda$, allowing the derivative to be expressed as
\begin{equation}
	-i \partial_\lambda \GFunc |_{\lambda =0 } =-i \partial_\lambda e^{i\phi(\lambda)} |_{\lambda =0} = \partial_\lambda \phi(\lambda)|_{\lambda=0} \approx \frac{\phi( \lambda) - \phi(0)}{ \lambda} = \frac{\phi( \lambda) }{ \lambda}
    \label{eq:limit}
\end{equation}
where in the last steps we have used the fact that, for symmetry reason, $\phi(0) = 0$ (see, for example, Ref. \cite{Solinas_2023}).

Then, the phase accumulated by the detector during the interaction with the system might be measured with interferometric techniques.
The other logical operations we need are the Hadamard gate $H$ and a phase gate $S$ defined by \cite{nielsen-chuang_book}
\begin{align}
H &= \frac{1}{\sqrt{2}} \begin{pmatrix}
1 & 1 \\
1 & -1
\end{pmatrix} \\
S &= \begin{pmatrix}
1 & 0 \\
0 & e^{i\frac{\pi}{2}}
\end{pmatrix} = \begin{pmatrix}
1 & 0 \\
0 & i
\end{pmatrix}
\end{align}
In terms of circuit implementation, the interferometric measure is built by applying a Hadamard gate, a phase gate, and another Hadamard to the detector and to measuring the populations of the two possible outcomes $P_{0,1}$ (see Fig. \ref{fig:algo_str}).

It can be shown (see Ref. \cite{Solinas_2023}) that, for small $\lambda$,  $\phi(\lambda) = -\arcsin(2P_0-1)$. Furthermore, using Eq. (\ref{eq:limit}) the derivative of the cost function $f(\vec \theta)$ reads
\begin{equation}
    g_l \sim -\frac{\arcsin(2P_0-1)}{ 2 \lambda\sin s}\,,
    \label{eq:QNDM_der_2}
\end{equation}
where we have added the term $2\sin s$ to the denominator to directly get the value of the derivatives according to the parameter shift rule. 


\section{Mean Squared Error and cost analysis}
\label{sec:bv}
Understanding how statistical errors affect the estimate of the derivatives and gradient plays a pivotal role in judging the performance and reliability of the different methods presented above. 
As a measure of the quality of the estimation of the cost function derivative, we consider Mean Squared Error ($MSE$), i.e., the average of the difference between the estimated values $\hat{g}_l$ and the exact value $g_l$, squared. 
We introduce the bias and variance as \cite{CoxHink74}
\begin{align}
    &\text{Bias}(\hat{g}_l)= \mathbb{E}(\hat{g}_l) - g_l \label{eq:Bias}\,,\\
    &\text{Var}(\hat{g}_l) = \mathbb{E}[(\hat{g}_l - \mathbb{E}(\hat{g}_l))^2]\,,\label{eq:var}
\end{align}
where $\mathbb{E}(\cdot)$ is the expectation over the statistical distribution of the measurement outcomes.
The bias represents a systematic distortion of a statistical result, and it remains present even in the limit of an infinite number of shots $N$. This is a consequence of the linear approximation used in Eq. \eqref{eq:limit}.
On the other hand, variance quantifies the degree of dispersion, indicating how far a collection of values deviates from its mean value. Unlike the bias, the variance depends on the shot number $N$ \cite{Mari2021}.
In terms of these quantities, we define the mean square error (MSE) as   
\begin{equation}
    \text{MSE}(\hat{g}_l)= \text{Bias}(\hat{g}_l)^2 + \text{Var}(\hat{g}_l)\,.
    \label{eq:MSE_teo}
\end{equation}

We start our discussion by computing the bias and variance for the \QNDM~method, following the methodology outlined in Ref. \cite{Mari2021}. They read

\begin{align}
    &\text{Bias}_{QNDM}(\hat{g}_l)= \left(\frac{\phi( \lambda)}{ \lambda} -i\partial_\lambda \GFunc\right) =\frac{\lambda \partial^2_\lambda \GFunc}{2} + \mathcal{O}(\lambda^2)\,, \label{eq:bias_QNDM}\\
    &\text{Var}_{QNDM}(\hat{g}_l) = \sigma_{QNDM}^2\,.
    \label{eq:var_QNDM}
\end{align}
We observe that the \QNDM~approach, as expected, exhibits bias different from zero arising from the linear approximation employed in Eq. (\ref{eq:limit}) to obtain the derivative of the quasi-characteristic function $\GFunc$. Therefore, its value is directly proportional to the second derivative of the function $\GFunc$ multiplied by $\lambda$. 
To compute the variance $\sigma^2_{QNDM}$,
we must incorporate the statistical error $\sigma^2_{P_0}$ associated with the detector population $P_0$ that we measure. 
For $N_{QNDM}$ single qubit measurements of the detector, we obtain $\sigma^2_{P_0}=\sigma^2_D/N_{QNDM}$
where $\sigma_D$ is the single shot variance obtained measuring the detector. Following the standard error propagation formulas \cite{taylor1997introduction} for Eq. (\ref{eq:QNDM_der_2}), we obtain $\sigma^2_{QNDM}$ from the variance for the detector population, which is given by
\begin{equation}
	\sigma^2_{QNDM}=\frac{\sigma^2_{P_0}}{4\sin^2 s} \frac{1}{\lambda^2(1-(2P_0-1)^2)}=\frac{\sigma^2_D}{4 N_{QNDM} \sin^2 s} \frac{1}{\lambda^2(1-(2P_0-1)^2)}.
    \label{eq:error_form_QNDM}
\end{equation}
Thus, the corresponding MSE is
\begin{equation}
     \text{MSE}_{QNDM}(\hat{g}_l)=\frac{\lambda^2 (\partial^2_\lambda \GFunc)^2}{4}+\frac{\sigma^2_D}{4 N_{QNDM} \sin^2 s}\frac{1}{\lambda^2(1-(2P_0-1)^2)}.
    \label{eq:MSE_QNDM}
\end{equation}

Notably, in MSE$_{QNDM}$, the variance term is directly proportional to $\lambda^{-2}$ while the bias term is proportional to $\lambda^2$.
Thus, the optimal reduction of the MSE hinges on selecting an appropriate value for $\lambda$. 

As highlighted in the previous section, the derivative of the quasi-characteristic can be approximately calculated if the coefficient in the argument of the system-detector coupling operator, $U_\pm = \exp\{\pm i \lambda \hat{Z_a} \otimes \hat{M} \}$, is small. Consequently, the appropriate choice of $\lambda$ depends, also, from the coefficients $h_i$ of observable $\hat{M} =  \sum_{i=1}^J h_i \hat{P}_i$. In the computational analysis, section \ref{sec:comp}, we present a significant example where a suitable value of $\lambda$ is selected.

In the \DM~approach, we need to measure the system qubits twice to obtain  $f(\vec \theta +s \hat{e}_l)$ and $f(\vec \theta -s \hat{e}_l)$. 
Proceeding as above, we have 
\begin{align}
    &\text{Bias}_{DM}(\hat{g}_l)= 0 \label{eq:bias_DM} \\
    &\text{Var}_{DM}(\hat{g}_l) = \frac{\sigma_{DM}^2(\vec \theta +s \hat{e}_l)+\sigma_{DM}^2(\vec \theta -s \hat{e}_l)}{4\sin^2s} \label{eq:var_DM}.
\end{align}

The bias contribution for \DM~vanishes because the parameter shift rule calculates the exact value of derivative \cite{Mari2021}. 
The variance, instead, is composed by the terms $\sigma_{DM}^2(\vec \theta \pm s \hat{e}_l)$. Intuitively, they correspond to the evaluation of $f(\vec \theta \pm s \hat{e}_l)$. 
Following Ref. \cite{Mari2021}, we assume that the variance of the measured observable depends weakly on the parameter shift such that $\sigma_{DM}^2(\vec \theta + s \hat{e}_l)+\sigma_{DM}^2(\vec \theta -s \hat{e}_l)\sim 2\sigma_{DM}^2$ for any value of $s$.
As we have seen in sec \ref{sec:DM}, to obtain the derivatives, we have to measure each Pauli string individually in the parameter space point $\vec \theta \pm s \hat{e}_l$ (see Eq. (\ref{eq:DM}) and \cite{Mari2021, Cerezo2021}). Then, the statistical error formula for the estimation of $f(\vec \theta)$ is given by
\begin{equation}
	 \sigma^2_{DM} = \sum_{i=1}^J  \frac{h^2_i\sigma^2_s}{N_{DM}}
    \label{eq:error_form_DM}
\end{equation}
where $h_i$ are the parameters in $\hat{M}$ in Eq. (\ref{eq:M}) and $\sigma_s$ is the single shot variance obtained measuring the $n$ qubits \cite{Mari2021}. In Eq. (\ref{eq:error_form_DM}), as in Ref. \cite{McClean2018, Mari2021}, we assume that $\sigma_s$ is equal for each Pauli string measurement.
Following these observations, we conclude that the MSE of \DM~approach is only due to the statistical noise and it is equal to
\begin{equation}
     \text{MSE}_{DM}(\hat{g}_l)= \sum_{i=1}^J \frac{h^2_i \sigma_s^2}{2 N_{DM} \sin^2s}.
     \label{eq:MSE_DM}
\end{equation}
A remark must be made regarding the results obtained in Eqs. (\ref{eq:MSE_QNDM}) and (\ref{eq:MSE_DM}). 
As demonstrated in section \ref{sec:comp}, an appropriate choice of $\lambda$ could provide a constant advantage in terms of error, giving to \QNDM~an advantage in terms of efficiency cost. 
\begin{equation}
   \frac{\text{MSE}_{DM}(\hat{g}_l)}{ \text{MSE}_{QNDM}(\hat{g}_l)}>1 
    \label{eq:N_DM}
\end{equation}

\subsection{Cost analysis and comparison between \DM~and \QNDM~approaches}
\label{sec:cost}

To make a comparison between the two approaches, we use the cost $\mathcal{C}\left(g_l^{i}\right)$ ($i=QNDM,\,DM$) that gives the scaling in terms of the total number of gates required to compute a derivative in the cost function for the two algorithms represented in Figs.  \ref{fig:algo_str_DM} and \ref{fig:algo_str}. In addition, we use the resource ratio $\mathcal{C}\left( g_l^{DM}/g_l^{QNDM}\right) = \mathcal{C}\left(g_l^{DM}\right) / \mathcal{C}\left(g_l^{QNDM}\right)$.

The resource costs express the number of gates, both single-qubit and two-qubit gates, used in one of two approaches. They are functions of the number of Pauli strings $J$, the number of qubits $n$, and the number of logical operators $k$ used in the unitary transformation $U(\theta)$, Eq. \eqref{eq:U_def}. As in the previous section, $N_{DM}$ and $N_{QNDM}$ represent the number of shots or repetitions for the two approaches. 

For \DM~approach, described in sec. \ref{sec:DM}, a circuit with $n$ qubits is transformed by a unitary operator $U(\theta - s e_l)$, and $P_i$ is measured $N_{DM}$ times. This process is repeated for each of the $J$ Pauli strings in $\hat{M}$ to calculate the expectation value $\langle M(\theta - s e_l) \rangle$. The same procedure is then repeated with the unitary operator $U(\theta + s e_l)$ to obtain $\langle M(\theta + s e_l) \rangle$. 
Summing all the contributions we get a computational cost equal to $2N_{DM}J(k + n)$.
For \QNDM~approach, described in sec. \ref{sec:QNDM}, a circuit with $n+1$ qubits is initially transformed by the operator $U_1 = U(\theta - s e_l)$, followed by the application of the exponential operator $U_- = \exp\{- i \lambda Z_a \otimes \hat{M} \}$. Next, the operator $U_2 = U^\dagger(\theta - s e_l) U(\theta - s e_l)$ is applied. Finally, the second exponential operator $U_+ = \exp\{+ i \lambda Z_a \otimes \hat{M} \}$ is applied, and the detector phase is measured. This procedure is repeated $N_{QNDM}$ times. The resulting computational cost is $N_{QNDM}(3k+8Jn)$.
The computed values of the resource costs are summarized in Tab. \ref{tab:cost_QNDM_DM}.

\begin{table}[h!]
\centering
\begin{tabular}{|c|c|c|}
\hline
\textbf{Method}& \textbf{Shots} & \textbf{Costs} \\
\hline
\DM & $N_{DM}$ & $\mathcal{C}\left(g_l^{DM}\right)=2N_{DM}J(k+n) $ \\
\hline
\QNDM & $N_{QNDM}$ & $ \mathcal{C}\left(g_l^{QNDM}\right)=N_{QNDM}(3k+8Jn)$\\
\hline
\end{tabular}
\caption{Resource cost function to compute first-order derivatives with the \DM~and the \QNDM~approaches. As can be seen, the \QNDM~has a clear cost advantage. However is import to emphasize that in the \QNDM~approach we run a more deeper quantum circuit, Fig. \ref{fig:algo_str} , than \DM, Fig. \ref{fig:algo_str_DM}.}
\label{tab:cost_QNDM_DM}
\end{table}
In this analysis, we focus on two regimes. The first occurs when $k \gg n J$, meaning that the number of logical gates required to implement $U(\vec \theta)$ exceeds the number of Pauli strings $J$ (multiplied by the number of qubits).
From Tab. \ref{tab:cost_QNDM_DM} and Ref. \cite{Solinas_2023}, we have that the resource ratio scales as
\begin{equation}
\mathcal{C}\left(\frac{g_l^{DM}}{g_l^{QNDM}}\right)= \frac{2J}{3k}\frac{N_{DM}}{N_{QNDM}} =\mathcal{O}(J).
\label{eq:k_gg_nJ}
\end{equation}

This result indicates that the cost of computing a derivative using \DM~is $J$ times higher than \QNDM~, leading to a linear speedup in this regime as the number of Pauli strings increases.
This regime is particularly relevant for quantum chemistry simulations. 
For instance, in simulations of moderately complex molecules \cite{Wecker2015}, typical values are  $k\approx 10^9-10^{10}$ operations for implementing $U(\vec \theta)$, $n \approx 10^2-10^3$ (corresponding to the expected size of near-term quantum processors), and $J > 10^3$ as the number of Pauli strings in the Hamiltonian. Given these parameters, the condition $k \gg n J$ holds, resulting in a reduction of approximately $J$ in resource usage when employing the \QNDM~approach. This advantage is expected to grow as simulations scale to larger and more complex molecules.

This regime holds also significant potential for simulations in quantum machine learning since the expressivity, i.e., the ability of a quantum circuit to generate (pure) states that are well representative of the Hilbert space \cite{Sim2019} of a quantum circuit is directly linked to the number of parametrized rotation in $U(\vec \theta )$, i.e., $k$.

The opposite regime is reached when $k \ll n J$, i.e., when the $\hat{M}$ operator is composed of many Pauli strings (thus, its averages might be difficult to estimate or measure) and the logical space to be searched is reduced (thus, we need unitary transformation with a limited number of logical gates).
In this case, the ratio between the resources employed reads
\begin{equation}
\mathcal{C}\left(\frac{g_l^{DM}}{g_l^{QNDM}}\right)=  \frac{k}{4n}\frac{N_{DM}}{N_{QNDM}} =\mathcal{O}(k).
\label{eq:k_ll_nJ}
\end{equation}
Whereas the $k$ factor comes from the linear cost advantage of \QNDM~with respect to \DM~for this regime (see Tab. \ref{tab:cost_QNDM_DM_2}).
This scenario is promising for various applications in VQAs as pointed out in \cite{Flynn_2022}.
\section{Computational analysis}
\label{sec:comp}
\subsection{Errors Analysis}
\label{sec:comp_mse}
In this section, we present a numerical comparison of error estimates obtained using the two methods. To gain a comprehensive understanding of the errors, we conducted multiple tests across different configurations. Here, we present one such test, where the number of qubits is $n=10$ and the Pauli strings in $\hat{M}$ are randomly selected from the set of all possible Pauli string operators. The coefficients $h_i$ in Eq. \eqref{eq:M} are drawn from a Gaussian distribution centered at $0$ with varying standard deviations. In all simulations, the results remain consistent with those obtained for a standard deviation of $5$, which we report here.
The analysis is performed as a function of the number of Pauli strings $J$.
As discussed in Sec. \ref{sec:bv}, to approximate the derivative of the $\GFunc$, the value of $\lambda$ must be taken small enough to justify the linear approximation in Eq. \eqref{eq:limit}.
As a consequence, the value of $\lambda$ depends (in a complex way) on the coefficients of the observable $\hat{M}$.
For this case, we set a value of $\lambda$ equal to the square root of the inverse of the sum of the absolute values of all coefficients $h_i$,
$\lambda = 1/\sqrt{\sum_i |h_i|}$.
For the numerical analysis, we have used the simulator provided by IBM Quantum known as Aer \cite{IBM_sim}. All the simulations are run to a shot count of $500$, a standard value for optimization.\\


The value of the derivatives $g_l$ depends on many parameters:
\begin{enumerate}
    \item[(j)] the logical gates needed to implement $U(\vec \theta)$, e.g., see Eq. (\ref{eq:U_tot});
    \item[(jj)] their total number $k$;
    \item[(jjj)] the direction $l$ along which is calculated the derivative;
    \item[(jv)] the Pauli strings $\hat{P}_i$ that appear in $\hat{M}$;
    \item[(v)] their number $J$, as pointed out in Eq. (\ref{eq:M}).
\end{enumerate}
For this reason, to ensure statistical significance, we need to average over these parameters.
In the following, we use the term “realization” to denote a random choice of all the {\it relevant} parameters and possibilities described in the points from $(j)$ to $(jv)$. 

\begin{figure}[h!]
    \centering
    \includegraphics[width=0.5\linewidth]{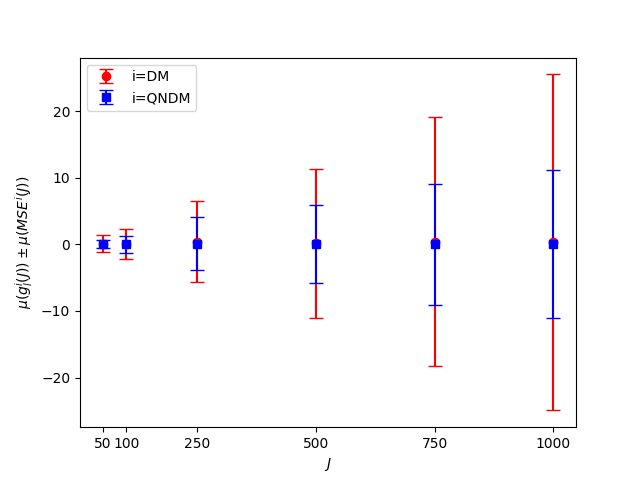}
    \caption{The plot represents the average derivative $\mu(g_l^i(J))$ along with the average mean squared error $\mu(MSE^i(J))$ [$i=$\DM~(red markers) and $i=$\QNDM~(blue markers)]. Each point is a function of the number of Pauli strings $J$. The results correspond to a shot count of $500$. The simulated quantum system consists of $n=10$ qubits. Each point is averaged over $L=100$ realizations, as discussed in the main text.}
    \label{fig:err_comp}
\end{figure}
In Fig. \ref{fig:err_comp}, we plot the derivative average $\mu(g_l^i(J))$ on the vertical axis, along with the average mean squared error $\mu(MSE^i(J))$. For each point, we calculate $L=100$ derivatives $g_l^i(J)$, varying the above parameters $(j)$ to $(v)$, and calculate their corresponding $MSE(g_l^i(J))$. Finally, we average all the derivatives and their corresponding MSEs.
\begin{equation}
    \mu(g_l^i(J)) = \frac{1}{L}\sum^{L=100}_{\Omega} g_l^i(J), \quad \mu(MSE^i(J)) = \frac{1}{L}\sum^{L=100}_{\Omega} MSE(g_l^i(J)),
\end{equation}
where the sum over $\Omega$ is intended over all the indices $(j),(jj),(jjj),(jv),(v)$.
The horizontal axis represents the number of Pauli strings $J$.
The \DM~and \QNDM~results correspond to the red and blue markers, respectively.

Notice that although the ratio between the errors of the \DM~and \QNDM~approaches remains constant as the number of Pauli strings increases, the \QNDM~approach with the same number of shots provides a growing advantage in terms of $MSE$. This means that, with the \DM~approach we must increase the number of shots to reach the same precision obtained with the \QNDM~one.

\subsection{Costs Comparison}
\label{sec:cost_sim}

For practical situations, it is of paramount importance to quantitatively estimate any advantage of the \QNDM~algorithm over \DM~and in which conditions this is reached.
From the discussion in Sec. \ref{sec:cost}, we expect the \QNDM~approach to perform better, i.e., it needs fewer resources, in some specific but interesting regimes. See for example Eqs. (\ref{eq:k_gg_nJ}) and (\ref{eq:k_ll_nJ}).
Aiming in this direction, we present a full numerical analysis of the resources needed to calculate the derivatives of the cost function. 
For a meaningful comparison, we fix the number of shots for the \QNDM~approach is set to $N_{QNDM} = 500$, while for the \DM~approach,  the number of shots is adjusted for each instance to ensure $MSE_{QNDM} = MSE_{DM}$. This means that if the MSE of \QNDM~for 500 shots is lower than the MSE of \DM~with 500 shots, the number of shots for \DM~must be increased (or vice versa) to satisfy the condition $MSE_{QNDM} = MSE_{DM}$.
As in the previous section, we present our cost analysis for a general example in which all coefficients of the observable $\hat{M}$ are drawn from a Gaussian distribution centered at $0$ with a standard deviation of $5$. As before, we took a value of $\lambda$ equal to the square root of the inverse of the sum of the absolute values of all coefficients $h_i$, $\lambda = 1/\sqrt{\sum_i |h_i|}$.

\begin{figure}[h!]
  \centering
  \includegraphics[width=1\linewidth]{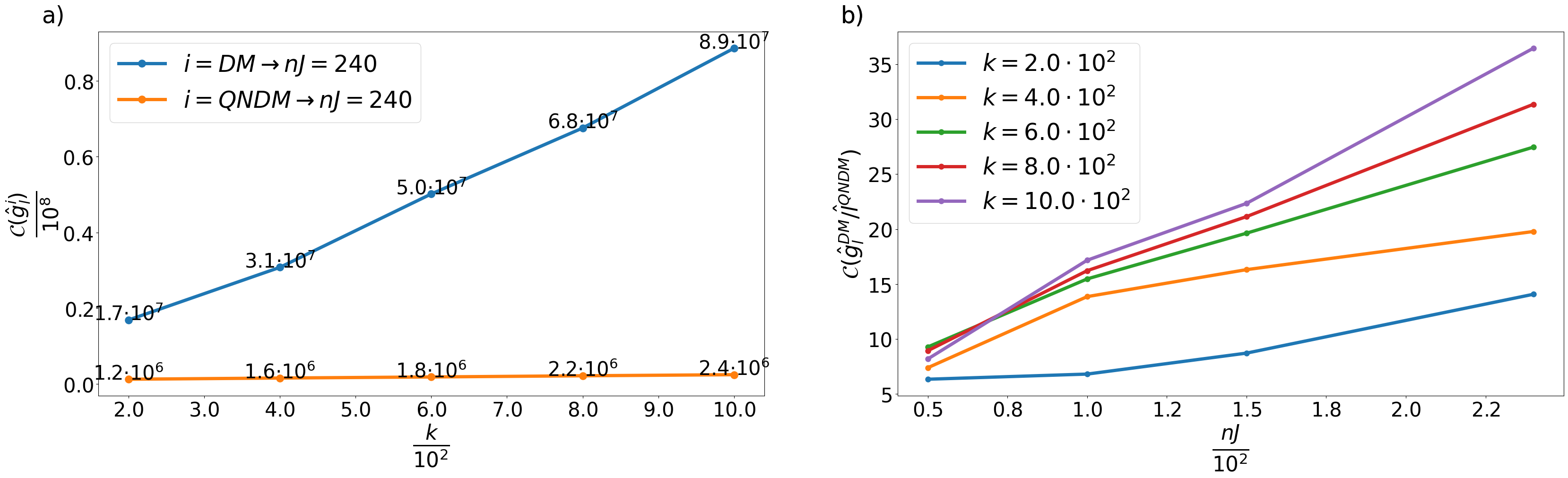} 
  \caption{Regime $k \gg nJ$.
  $a)$ The averaged number of resources $\mathcal{C}(g_l^i)$ needed to estimate the derivative of the cost function along the direction $l$; here, $i=QNDM$ (orange line) and $i=DM$ (blue line). $\mathcal{C}(g_l^i)$ is plotted as a function of the number of logical operators $k$ for a fixed $n J=240$. 
  $b)$ Ratio between \DM~and \QNDM~resources numbers $\left(\mathcal{C}\left(g_l^{DM}/g_l^{QNDM}\right)\right)$ as a function of Pauli string $J$ each colored line corresponds at a fixed $k$. The quantum circuit is composed of $n=10$ qubits and the average is taken over $L=50$ realization as discussed in the main text. The number of shots for the \QNDM~is $N_{QNDM}=500$, for \DM, it is calculated for each realization such that the MSE errors are equal for both methods. 
  }
  \label{fig:cost_q_5_w}
\end{figure}

The results of Fig. \ref{fig:cost_q_5_w} illustrate the expected advantage of \QNDM~over the \DM~approach.
Panel a), on the left, shows the resource cost $\mathcal{C}\left(g_l^{i}\right)$ for the two approaches in the limit $k \gg n J$. They are plotted as a function of the number of logical operators $k$ needed to implement $U(\vec{\theta})$. The simulations have been done for $n=10$ qubits and a with $J=24$ different Pauli strings.
Each value in Fig. \ref{fig:cost_q_5_w} is calculated by averaging over different realizations with randomly chosen values of the parameters $J$, $k$, and the set of Pauli strings in $\hat M$.
The average is performed also over the unitary transformations $U(\vec{\theta})$. 
This is obtained through a series of layers parameterized by $U_j(\vec \theta_j)$ and entanglement layers denoted as $V_j$,  as illustrated in Eqs. (\ref{eq:rot_layer}) and (\ref{eq:ent_layer}). Within the parameterized layer, the constituent single-qubit rotations $R^j_i(\theta_i^j)$, and their corresponding parameters $\vec \theta_i^j$, are selected at random.

In panel a) of Fig. \ref{fig:cost_q_5_w}, as expected, \QNDM~needs fewer resources than \DM~to evaluate the derivatives.
Less obvious is that already for a limited number of logical operators, i.e. $k\sim 10^2$, we achieve a reduction of more than an order of magnitude in the number of logical gates.

In panel b) of Fig. \ref{fig:cost_q_5_w}, on the right, we show the ratio of the resources in terms of the number of Pauli strings $J$, comparing different possible choices for $k$.
Already for $k \sim 10^2$, the ratio closely follows the theoretical linearity predicted in Eq. (\ref{eq:k_gg_nJ}). In this regime, the \DM~method requires more than 35 times the resources of the \QNDM~approach.

\begin{figure}[h!]
  \centering  \includegraphics[width=1\linewidth]{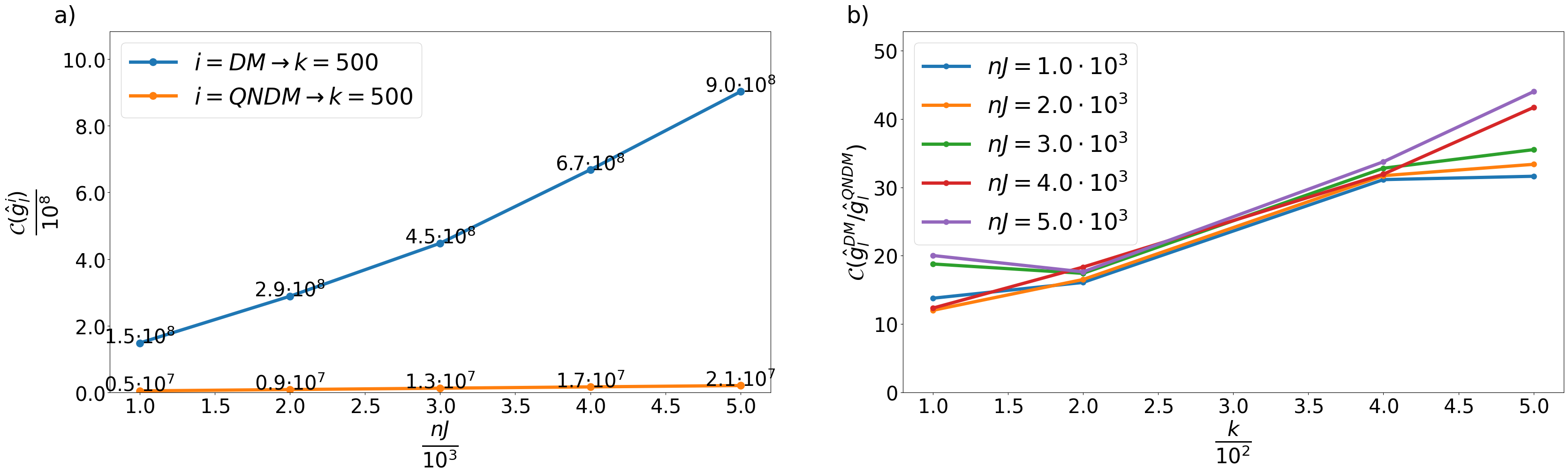}
  \caption{
  Regime $k \ll n J$. The averaged number of resources $\mathcal{C}(g_j^i)$ needed to estimate the derivative of the cost function along the direction $j$ ($i=QNDM$(orange line) and $i=DM$(blue line)).
 $\mathcal{C}(g_j^i)$ is plotted as a function of the product $n J$ for a fixed number of logical operators $k=0.5 \times 10^3$. $b)$ Ratio between \DM~and \QNDM~resources $\left(\mathcal{C}\left(g_j^{DM}/g_j^{QNDM}\right)\right)$ as a function of the number of logical operators $k$ each colored line corresponds at a fixed $n J$. The quantum circuit is composed of $n=10$ qubits and the average is taken over $L=50$ realization as discussed in the main text. The number of shots for the \QNDM is $N_{QNDM}=500$, for \DM, it is calculated for each realization such that the MSE errors are equal for both methods.
}
  \label{fig:cost_q_5_h}
\end{figure}

The regime in which $k \ll n J$ is analyzed in Fig. \ref{fig:cost_q_5_h}. Panel a), on the left, shows the resource cost function $\mathcal{C}(g_l^i)$ ($i=QNDM, DM$) obtained to estimate the derivative along the direction $l$. The values are plotted as a function $n J$. The simulations are done for $n=10$ qubits with $k=0.5\cdot10^3$ fixed. Each value in Fig. \ref{fig:cost_q_5_h} is calculated by performing an average over a collection of different realizations, as discussed above.
As delineated in tab. \ref{tab:cost_QNDM_DM}, for \DM~(the blue line) the number of resources increases linearly in the number of $J$.
Already for $nJ \sim 10^3$, the \DM~method already requires 10 times more resources than the \QNDM~one.

In panel b), on the right, we show the ratio of the resources as a function of the number of logical operators $k$ and for different values of $n J$.
The expected advantage, Eq. (\ref{eq:k_ll_nJ}), of the \QNDM~approach is confirmed in this regime. Already for $n J \sim 10^3$ the \DM~approach needs $40$ times more resources than the \QNDM~one.

Notice in all our simulations and both regimes, we have considered relatively small and simple systems to provide a more conservative estimate. We might expect a further increasing advantage in realistic computational problems, as they usually require more logical gates and involve more complex observables $\hat{M}$.



\section{Second derivative and Hessian calculation}
\label{sec:second_derivative}
The \QNDM~approach can be naturally extended to the calculation of higher derivatives \cite{Solinas_2023}.
Having access to the second derivatives of the cost function allows us to evaluate the Hessian matrix. This can be useful for the implementation of second-order optimizers like the Newton optimizer or the Diagonal Newton optimizer \cite{Mari2021}.

With a straightforward extension of Eq. (\ref{eq:f_grad_def}), it can be shown that the second derivative, requires the evaluation of the cost function $f$ in four points \cite{Mari2021, Solinas_2023}.
To implement this in the \DM~approach, we need four independent measurements of the observables $\hat{M}$.
On the other hand, following the \QNDM~paradigm, we can just couple the system to the detector four times to store in the phase of the latest the same amount of information \cite{Solinas_2023}. Then, analogously to the first-order derivative, we perform a measure on the detector only once.  
For further details and the numerical simulations about the second derivatives with \QNDM~method, we address the interested reader to Appendix \ref{sec:higher_derivatives}. In this section, instead, we present the persistence of the advantages observed in first derivatives even when extending to second derivatives.

\subsection{Costs Analysis for second derivatives}
\label{sec:second}
In Table \ref{tab:cost_QNDM_DM_2}, we insert the costs for calculating second derivatives. We denote the resource cost function, $\mathcal{C}\left(g_{w,l}^{i}\right)$, for both the approaches ($i=QNDM,\,DM$) in terms of the number of gates required to compute the second derivative along the directions $w$ and $l$. 
These values are obtained from Refs. \cite{Solinas_2023, Cerezo2021, Mari2021}. As above, the costs depend on the number of Pauli strings $J$, the number of qubits $n$, the number of shots or repetitions $N_i$, and the number of logical operators needed to implement $U(\vec \theta)$, denoted by $k$. 

\begin{table}[h!]
\centering
\begin{tabular}{|c|c|c|}
\hline
\textbf{Method}& \textbf{Shots} & \textbf{Costs} \\
\hline
\DM & $N_{DM}$ & $\mathcal{C}\left(g_{w,l}^{DM}\right)=N_{DM}4J(k+n) $ \\
\hline
\QNDM & $N_{QNDM}$ & $ \mathcal{C}\left(g_{w,l}^{QNDM}\right)=N_{QNDM}(7k+16Jn)$\\
\hline
\end{tabular}
\caption{Resource cost function to compute second derivatives with the \QNDM~and the \DM~approaches.}
\label{tab:cost_QNDM_DM_2}
\end{table}

From Table \ref{tab:cost_QNDM_DM_2}, we observe that the cost of calculating the second derivative with \DM~is twice the cost for the first derivative with the same method. For \QNDM, the variation in cost is due to the coefficient of $k$ (which increases from $3$ to $7$) and to the coefficient of $J$, which is twice the one for a single derivative. A detailed description of the implementation of both approaches for computing the second derivative is provided in Appendix \ref{sec:higher_derivatives}.

For a fixed value of MSE error and when $k \gg n J$, the ratio of the resource cost functions reads
\begin{equation}
\mathcal{C}\left(\frac{g_{w,l}^{DM}}{g_{w,l}^{QNDM}}\right)=\frac{4J}{7}\frac{N_{DM}}{N_{QNDM}} =\mathcal{O}(J)
\label{eq:k_gg_nJ_2}
\end{equation}
The opposite regime is $k \ll n J$. In this case, the resource ratio is given by
\begin{equation}
\mathcal{C}\left(\frac{g_{w,l}^{DM}}{g_{w,l}^{QNDM}}\right)=  \frac{k}{4n}\frac{N_{DM}}{N_{QNDM}} = \mathcal{O}(k)
\label{eq:k_ll_nJ_2}
\end{equation}
These results demonstrate that the advantage in terms of the cost function for \QNDM~is also maintained in the case of the second derivative. As already mentioned, all the simulations used to validate these results are included in Appendix \ref{sec:higher_derivatives}.

\section{Conclusions}
\label{sec:conclusions}

In this paper, we have presented a detailed study of an alternative approach to evaluating the derivatives of a cost function using a quantum computer, along with its implementation in $Python$ publicly available via GitHub \cite{qndm_gradient}.
This method called Quantum Non-Demolition Measurement (\QNDM), was first introduced in Ref. \cite{Solinas_2023}. In addition to providing a quantitative estimate of the resources needed to run the \QNDM~protocol, we have conducted an in-depth comparison with the state-of-the-art Direct Measurement (DM) approach \cite{Cerezo2021}.
The results obtained are encouraging and pave the way for the first applications of the \QNDM~ algorithm to real-world problems since it allows us to consistently reduce the resources needed for the computation. 

This advantage, as highlighted in \cite{Solinas_2023}, arises from the use of a quantum detector to store information about the derivative of the cost function, allowing us to run the quantum circuit and perform measurements only once. In contrast, the \DM~approach requires executing two separate circuits, with each circuit needing to be repeated for the number of Pauli strings. For the second derivative case, this requirement increases to four circuits.

The numerical simulations performed using the $Qiskit$ framework have confirmed the analytical estimates. Even for small systems and simplified quantum circuits, we observe a considerable reduction in the resources needed to run the \QNDM~approach compared to the \DM~method. We anticipate that this advantage will further increase for more practical and complex problems, such as the simulation of chemical compounds, molecules, or small quantum systems \cite{Weber2014, sawaya2024hamlib}.
These results position the \QNDM~approach as a valuable alternative for implementing Variational Quantum Algorithms (VQAs) on the next generation of noisy quantum computers.


\begin{acknowledgments}
PS, GM, and DM acknowledge financial support from INFN. This work was carried out while GM was enrolled in the Italian National Doctorate on Artificial Intelligence run by Sapienza University of Rome in collaboration with the Dept. of Informatics, Bioengineering, Robotics, and Systems Engineering, Polytechnic School of Genova.\\
SC would like to thank the University of Amsterdam and the Delta-ITP program for their support during the initial part of this work.
\end{acknowledgments}


\bibliographystyle{apsrev4-1}
\bibliography{qndm_comp}


\appendix
\setcounter{equation}{0}

\pagebreak
\widetext
\begin{center}
\textbf{\large Appendixes}
\end{center}
\setcounter{equation}{0}
\setcounter{figure}{0}
\setcounter{table}{0}
\setcounter{page}{1}
\makeatletter
\renewcommand{\theequation}{S\arabic{equation}}
\renewcommand{\thefigure}{S\arabic{figure}}
\renewcommand{\bibnumfmt}[1]{[#1]}
\renewcommand{\citenumfont}[1]{#1}

\section{Second derivatives}
\label{sec:higher_derivatives}
\begin{figure}[h!]
  \centering
  \includegraphics[width=0.7\linewidth]{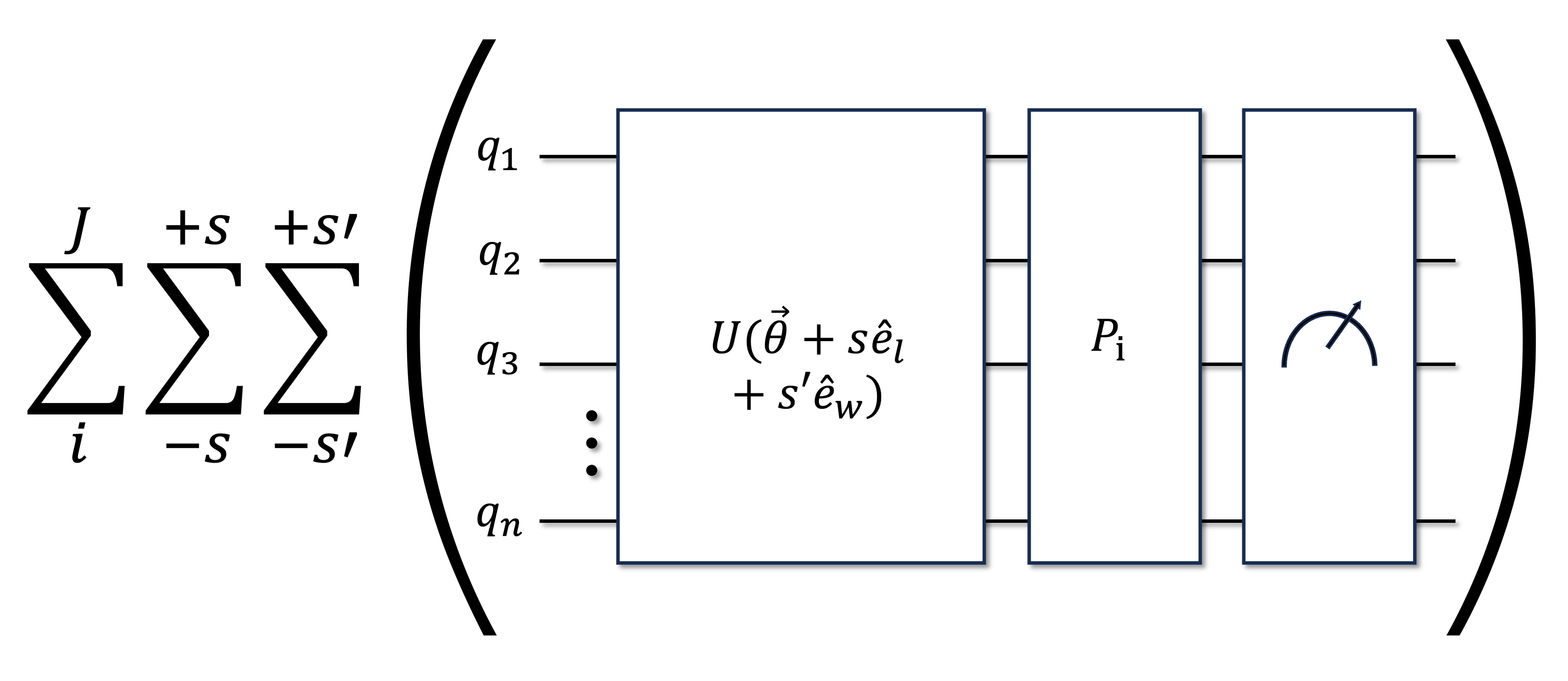}  
  \caption{Quantum circuit of the implementation of the \DM~protocol. Here $U(\vec \theta +s\hat{e}_l+s'\hat{e}_w)$ is Eq. (\ref{eq:U_def}) with two parameters shifted in directions $\hat{e}_l$ and $\hat{e}_w$. The $J$ is equal to the number of Pauli strings and the values $\pm s$ and $\pm s'$ are the shift for the parameter shift rule.}
  \label{fig:algo_str_DM_2}
\end{figure}

In this section, we present a detailed analysis of the second derivative. The parameter shift rule, Eq. (\ref{eq:f_def}), can be generalized to calculate the second derivative \cite{Mari2021}.
\begin{eqnarray}\label{eq:second_derivative}
	g_{w,l} &=& \frac{\partial^2 f({\vec \theta})}{\partial \theta_l \partial \theta_w} = \\ \nonumber 
	&=&  
	\Big[ f(\vec \theta +s (\hat e_l+\hat e_w)) - f(\vec \theta +s (-\hat e_l+\hat e_w)) \\ \nonumber 
	&-&f(\vec \theta +s (\hat e_l-\hat e_w)) + f(\vec \theta -s (\hat e_l+\hat e_w))\Big] \Big[2 \sin^2 s \Big]^{-1}.
    \label{eq:second_psr}
\end{eqnarray}
We can re-write Eq. (\ref{eq:second_psr}) in terms the density matrix representing the $n$-qubits initial state, i.e. $\rho_s^0=|\psi_0\rangle\langle\psi_0|$
\begin{eqnarray}
	\frac{\partial^2 f({\vec \theta})}{\partial \theta_l \partial \theta_i} &=& \Big[\Trace_S [ 
	U^\dagger (\vec \theta + s (e_l+e_w)) \hat M U (\vec \theta + s (e_l+e_w)) \rho_S^0 
	-U^\dagger (\vec \theta + s (-e_l+e_w)) \hat M U (\vec \theta + s (-e_l+e_w)) \rho_S^0 \nonumber \\	
	&-&U^\dagger (\vec \theta + s (e_l-e_w)) \hat M U (\vec \theta + s (e_l-e_w)) \rho_S^0 
	+U^\dagger (\vec \theta - s (e_l+e_w)) \hat M U (\vec \theta - s (e_l+e_w)) \rho_S^0 ]\Big]\Big[4 \sin^2 s \Big]^{-1}. \nonumber \\
    \label{eq:second_rho}
\end{eqnarray}
It follows that for \DM~method to calculate the values of $f(\vec \theta +s (\hat e_l+\hat e_w))$, we run the quantum circuit to implement the corresponding $U(\vec \theta +s (\hat e_l+\hat e_w))$ and then perform a projective measurement for each Pauli string composing the observable $\hat M$ \cite{Mari2021}.
The same procedure is then implemented for $f(\vec \theta +s (-\hat e_l+\hat e_w)), f(\vec \theta +s (\hat e_l-\hat e_w)), f(\vec \theta -s (\hat e_l+\hat e_w))$, then the derivative can be easily extracted. In Fig. \ref{fig:algo_str_DM_2} the \DM is represented schematically.

\begin{figure}[h!]
  \centering
  \includegraphics[width=0.8\linewidth]{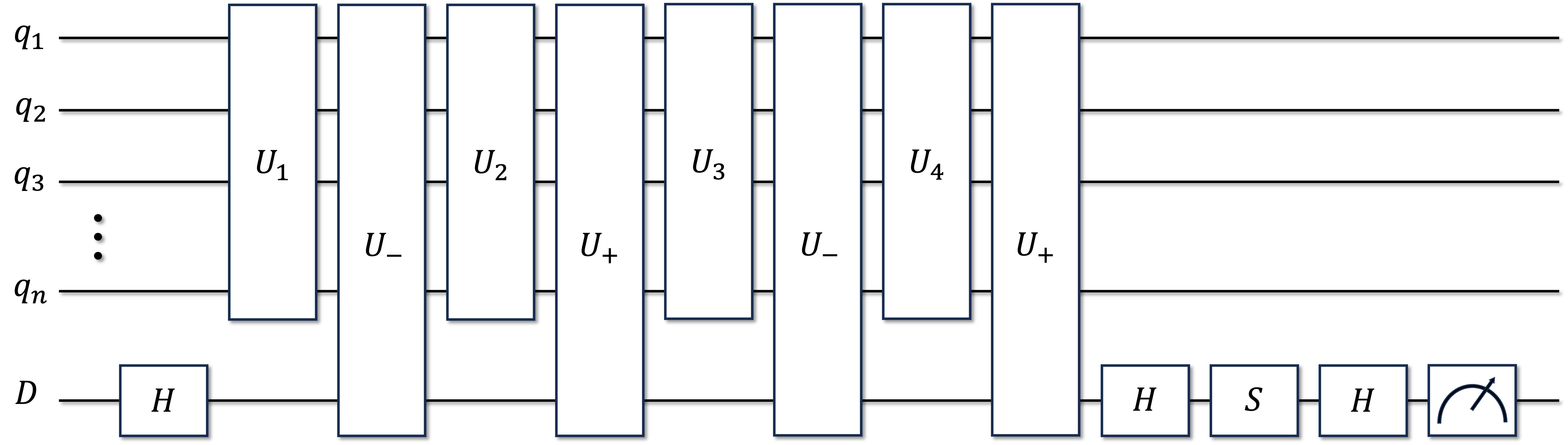}  
  \caption{Quantum circuit of the implementation of the \QNDM~protocol for second derivatives. 
  Here, $H$ is the Hadamard gate, $S$ is the phase gate, $U_1 = U (\vec \theta + s (e_l-e_w))$, 
  $U_2 =
	U^\dagger (\vec \theta + s (e_l-e_w))U (\vec \theta - s (e_l+e_w))$, 
 $U_3 = 
	U^\dagger (\vec \theta - s (e_l+e_w))U (\vec \theta + s (-e_l+e_w))$, 
 $U_4 = U^\dagger (\vec \theta + s (-e_l+e_w))U (\vec \theta + s (e_l+e_w))$ 
 and $U_\pm = \exp\{\pm i \lambda \hat{Z_a} \otimes \hat{M} \}$ is the system-detector coupling operator.}
  \label{fig:algo_str_2}
\end{figure}

On the other hand, in the \QNDM~method \cite{Solinas_2023}, we estimate the second derivative measuring the circuit shown in Fig. \ref{fig:algo_str_2}, and we avoid the repetitions for each Pauli string and for each term of the parameter shift rule.
The unitary transformation corresponding to the full \QNDM~evolution is 
\begin{equation}
    U^2_{tot} = e^{ i \lambda Z_a \otimes \hat{M} } U_4 e^{-i \lambda Z_a \otimes \hat{M} } U_3 e^{ i \lambda Z_a \otimes \hat{M} } U_2 e^{-i \lambda Z_a \otimes \hat{M} } U_1,
\end{equation}
where the $U_y$ operators (with $y=1,..,4$) correspond to the operators working in the $\vec \theta$ space and they read
\begin{eqnarray}
	U (\vec \theta + s (e_l-e_w)) &=& U_1 \nonumber \\
	U^\dagger (\vec \theta + s (e_l-e_w))U (\vec \theta - s (e_l+e_w)) &=& U_2 \nonumber \\
	U^\dagger (\vec \theta - s (e_l+e_w))U (\vec \theta + s (-e_l+e_w)) &=& U_3  \nonumber \\
	U^\dagger (\vec \theta + s (-e_l+e_w))U (\vec \theta + s (e_l+e_w)) &=& U_4.
\end{eqnarray}

\subsection{Errors Analysis for Second Derivative}
The MSEs for the second derivatives for both methods, respectively, are equal to
\begin{align}
      \text{MSE}_{QNDM}(\hat{g}_{w,l})&=\frac{\lambda^2 (\partial^2_\lambda \GFunc)^2}{4}+\frac{\sigma_D^2}{16N_{QNDM}\sin^4s}\frac{1}{\lambda^2(1-(2P_0-1)^2)}
     \label{eq:MSE_QNDM_2}
      \\
      \text{MSE}_{DM}(\hat{g}_{w,l})&= \sum^J_i= \frac{h^2_i\sigma_s^2}{4N_{DM}\sin^4s}.
     \label{eq:MSE_DM_2}
\end{align}

\subsection{Cost Simulations for Second Derivative}

\begin{figure}[h!]
  \centering
  \includegraphics[width=1\linewidth]{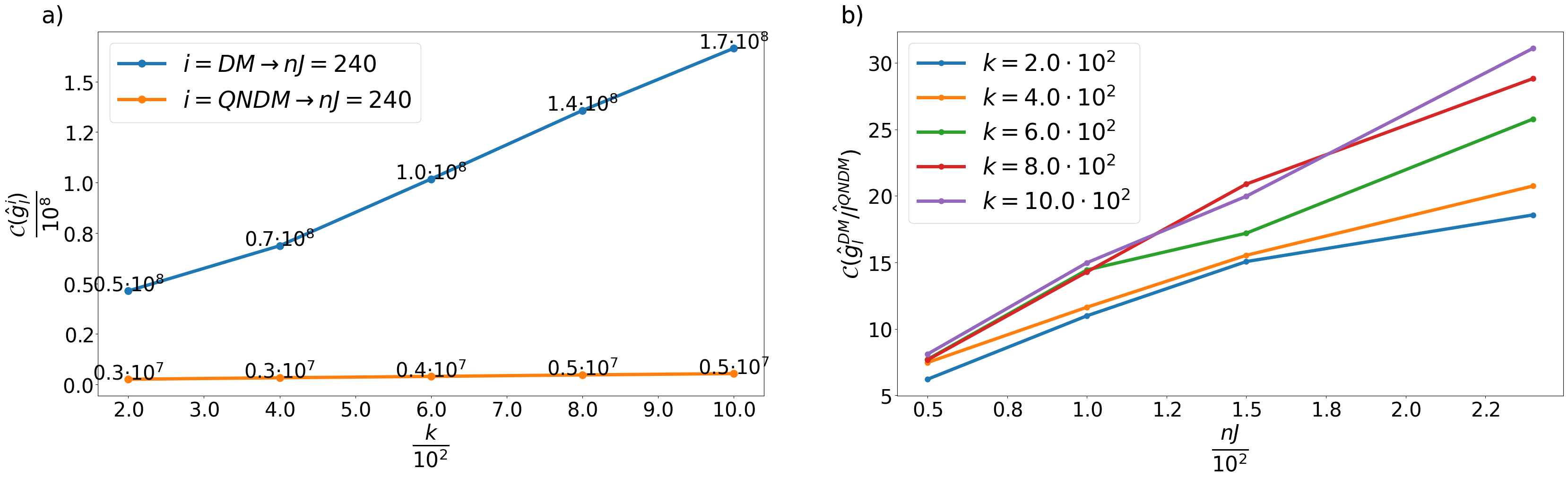} 
  \caption{Regime $k \gg nJ$.
  $a)$ Number of resources $\mathcal{C}(g_{w,l}^i)$ needed to estimate the second derivative of the cost function along the directions $l$ and $w$; here, $i=QNDM$ (orange line) and $i=DM$ (blue line). $\mathcal{C}(g_{w,l}^i)$ is plotted in function of the number of logical operators $k$ for a fixed $nJ=240$. 
  $b)$ Ratio between \DM~and \QNDM~resources numbers $\mathcal{C}\left(g_{w,l}^{DM}/g_{w,l}^{QNDM}\right)$ as a function of Pauli string $nJ$ each colored line corresponds at a fixed $k$. The quantum circuit is composed of $n=10$ qubits and the average is taken over $L=50$ realization as discussed in the main text. The number of shots for the \QNDM~is $N_{QNDM}=500$, for \DM, it is calculated for each realization such that the MSE errors are equal for both methods.}
  \label{fig:cost_q_5_w_2}
\end{figure}

\begin{figure}[h!]
  \centering
  \includegraphics[width=1\linewidth]{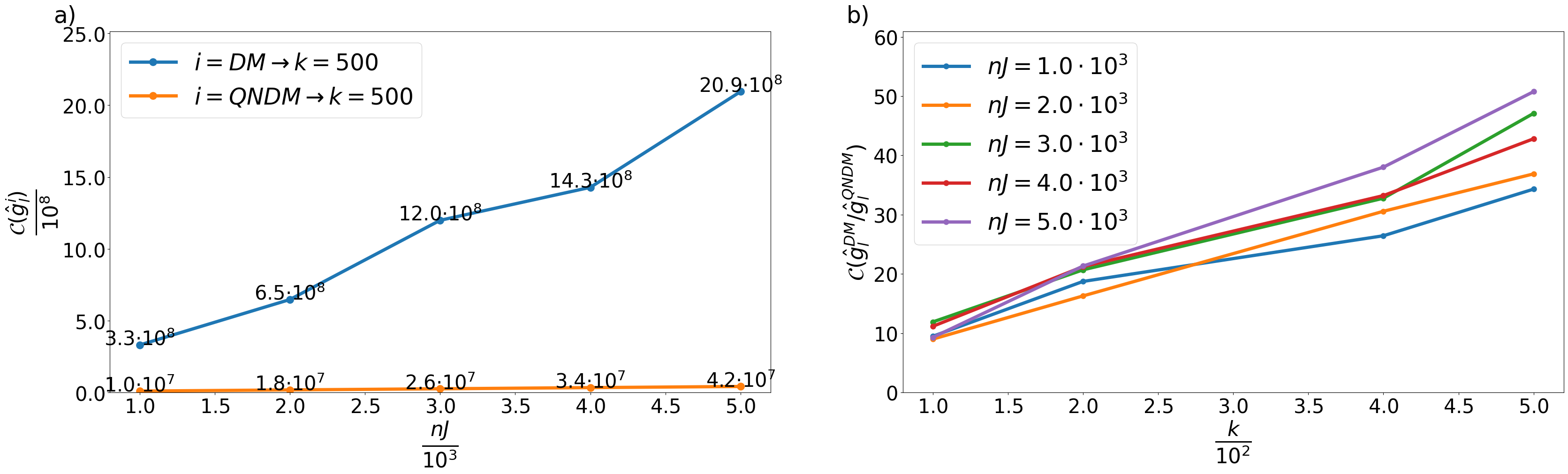}
  \caption{$a)$ The averaged number of resources $\mathcal{C}(g_{w,l}^i)$ needed to estimate the second derivative of the cost function along the directions $l$ and $w$; here, $i=QNDM$ (orange line) and $i=DM$ (blue line). $\mathcal{C}(g_{w,l}^i)$ is plotted in function of the number of Pauli strings $J$ for a fixed number of logical operators $k=5\cdot10^3$. $b)$ Ratio between \DM~and \QNDM~resources numbers $\mathcal{C}\left(g_{w,l}^{DM}/g_{w,l}^{QNDM}\right)$ as a function of logical operators $k$ each coloured line corresponds at a fixed $n J$. The quantum circuit is composed of $n=10$ qubits and the average is taken over $L=50$ realization as discussed in the main text. The number of shots for the \QNDM is $N_{QNDM}=500$, for \DM, it is calculated for each realization such that the MSE errors are equal for both methods.}
  \label{fig:cost_q_5_h_2}
\end{figure}

In Section \ref{sec:second}, we found the asymptotic cost ratios $\mathcal{C}\left( g_{w,l}^{DM}/g_{w,l}^{QNDM}\right)$ for the two regimes: $k\ll nJ$ Eq. (\ref{eq:k_ll_nJ_2}) and $k\gg nJ$ Eq. (\ref{eq:k_gg_nJ_2}). In Figs. \ref{fig:cost_q_5_h_2} and \ref{fig:cost_q_5_w_2}, we include simulations that illustrate how the theoretical results are also maintained for the second derivatives. For each value presented in the figures, we averaged over $L=50$ different realizations. 
Panel a) in Fig. \ref{fig:cost_q_5_w_2} shows the numbers of resources, $\mathcal{C}\left(g_{w,l}^{i}\right)$, in the function of the logical operators $k$, when we consider the limit of $k \gg n J$. The simulations are done for $n=10$ qubits, fixed $n J=240$, and fixed shots $N=500$.
In the second panel b) we show the ratio of the resources as a function of the number of Pauli strings $J$ and for $k$ different numbers of logical operators in $U(\vec \theta)$.
Panel a) in Fig. \ref{fig:cost_q_5_h_2} shows the numbers of resources, $\mathcal{C}\left(g_{w,l}^{i}\right)$, in function of $n J$. We are working in the limit $k \ll n J$. The simulations are done for $n=10$ qubits, a fixed number of logical operators $k=5\cdot10^2$, and fixed shots $N=500$.
In the second panel b) we show the ratio of the resources as a function of logical operators $k$ for different values of the product $n J$.
The advantage of the \QNDM~over the \DM~approach is confirmed also for the second derivatives.

\newpage
\section{how to install \QNDM}
\label{sec:git}

To install the \QNDM~library just clone the {\it GitHub} repository (link in \cite{qndm_gradient}) using the command
\vspace{\baselineskip}
\begin{center}
\fbox{
    \parbox{0.5\textwidth}{
    \centering
    git clone https://github.com/simonecaletti/qndm-gradient.git}
}
\end{center}
\vspace{\baselineskip}
All the functions for the \QNDM~algorithm are defined in the {\it qndm} folder. The {\it test\_scripts} folder instead contains a set of scripts with some simple tests, like \QNDM~and \DM~derivatives evaluation and optimization tasks.
To have access to the interface with the hamlib library \cite{sawaya2024hamlib} you need to install the {\it mat2qubit} package \cite{sawaya2022mat2qubit}. The instructions are contained in the {\it hdf5-install.sh} script, so just run 
\vspace{\baselineskip}
\begin{center}
\fbox{
    \parbox{0.5\textwidth}{
    \centering
    bash hdf5-install.sh}
}
\end{center}
\vspace{\baselineskip}
HamLib (for Hamiltonian Library), is freely available online and contains problem sizes ranging from 2 to 1000 qubits. It includes problem instances of the Heisenberg model, Fermi-Hubbard model, Bose-Hubbard model, molecular electronic structure, molecular vibrational structure, MaxCut, Max-k-SAT, Max-k-Cut, QMaxCut, and the traveling salesperson problem \cite{sawaya2024hamlib}.
To test the installation create the folder {\it output\_test} and run a test script, for example 
\vspace{\baselineskip}
\begin{center}
\fbox{
    \parbox{0.5\textwidth}{
    \centering
    python3 test\_qndm.py}
}
\end{center}
\vspace{\baselineskip}
Run it from the {\it qndm-gradient/} folder or add the corresponding path to your PYTHON PATH environment variable.
If the installation is working correctly a {\it QNDM\_der.csv} and a {\it RunCard\_Der.txt} file have been created. The first one contains information about the gradient computation using the \QNDM~algorithm, while the second is an automatically generated runcard containing the details of the run.

\end{document}